# Structural origins of ultra-low glass-like thermal conductivity in AgGaGe$_3$Se$_8$

Peter Skjøtt Thorup,[a] Rasmus Stubkjær Christensen,[a] Kim-Khuong Huynh,[a] Pavankumar Ventrapati,[a] Emilie Skytte Vosegaard,[a] and Bo Brummerstedt Iversen[a*]

[a]Center for Integrated Materials Research, Department of Chemistry and iNANO, Aarhus University, DK-8000 Aarhus, Denmark

*Corresponding author: bo@chem.au.dk

**Abstract**

Materials with a low thermal conductivity are important for a variety of applications such as thermal barrier coatings and thermoelectrics, and understanding the underlying mechanisms of low heat transport, and relating them to structural features, remains a central goal within material science. Here, we report on the ultra-low thermal conductivity of the quarternary silver chalcogenide AgGaGe$_3$Se$_8$ with a remarkable value of only 0.2 Wm$^{-1}$K$^{-1}$ at room temperature, and an unusual glass-like thermal behavior from 2 K to 700 K. The ultra-low thermal conductivity is linked to a disordered nature of Ag in the structure, displaying extremely large Ag atomic displacement parameters obtained from multi-temperature synchrotron powder X-ray scattering measurements, and Ag ionic conductivity at elevated temperatures. Additionally, signs of structural anharmonicity and soft bonding are evident from a low temperature Boson peak in the heat capacity and a low Debye temperature of 147 K.

1. Introduction

Materials that exhibit low thermal conductivity are of great scientific and technological interest, and are crucial for thermal management of electronic devices, thermal barrier coatings, and in thermoelectric materials.[1-2] The mechanisms of heat transport in solids are very complex and depend strongly on the atomic structure of the materials with completely amorphous solids representing the lower limit. Besides free charge carriers, the main contribution to the thermal conductivity in crystalline materials is associated with lattice vibrations. The quantized modes of lattice vibrations are called phonons and using kinetic gas theory the lattice thermal conductivity ($\kappa_L$) can be described by $\kappa_L = \frac{1}{3} C_v v_g \lambda_{ph}$, where $C_v$ is the volumetric heat capacity, $v_g$ the phonon group velocity, and $\lambda_{ph}$ the phonon mean free path. Heat transported by phonons can be reduced by either lowering the mean free path through phonon scattering or lowering the phonon group velocity *e.g.* by soft chemical bonds or heavy atomic masses. The phonon scattering mechanisms can be further divided into intrinsic effects, such as phonon-phonon scattering that is affected by lattice anharmonicity, disorder, rattling dynamics etc., and extrinsic effects such as solid-solution defects, boundary scattering, and micro/nanostructures.

The lower limit to the thermal conductivity of materials has been discussed and investigated since Einstein's theoretical work of describing the atoms in a solid as independent harmonic oscillators.[3] Due to the wave characteristics of phonons there is a lower limit to the phonon scattering length of half the wavelength, often referred to as the Ioffe-Regel limit, where propagating phonons are scattered so frequently that the phonon picture breaks down. The resulting minimum thermal conductivity in this limit has been described by Cahill, Watson, and Pohl based on random hops of thermal energy between uncorrelated oscillators and is often referred to as the glass limit. [4] Another type of heat conduction can be described by the diffuson-mediated heat transport mechanism. Heat transport via the diffuson was described by Allen and Feldman[5] and is represented as non-propagating, non-localized heat-carrying vibrations at intermediate energies, and is proposed to be a better description of atomic vibrations in materials with more atomic disorder, including large complex unit cells.[6] Diffuson-mediated transport was first used to describe amorphous materials, however it has recently been shown that diffusons can be a significant part of heat transport in complex crystal structures such as $Yb_{14}MnSb_{11}$,[7] and can be associated with materials exhibiting fast ion transport.[8] The low thermal conductivity associated with fast ion conductors is reflected in many promising thermoelectric materials being based on mixed ionic-electronic conductors such as $Zn_4Sb_3$,[9] $Cu_2Se$,[10] and $Ag_9GaSe_6$.[11]

In the present paper, we report the ultra-low thermal conductivity for $AgGaGe_3Se_8$ with a glass-like temperature dependence between 2 K and 700 K and a value of 0.2 Wm$^{-1}$K$^{-1}$ at room temperature. This is extraordinarily low for a crystalline compound and lies between the calculated glass and the diffuson limit of minimum thermal conductivity. The quaternary compound of $AgGaGe_3Se_8$ has exclusively been studied for its non-linear optical properties in the literature,[12] and the thermal properties are unexplored. Here the structural features of the thermal properties are investigated through multi-temperature synchrotron powder X-ray diffraction (PXRD) experiments, which reveal anomalously high anisotropic atomic displacement parameters (ADPs) for Ag. These large Ag displacements corroborate with void space analysis and measurements of ionic conductivity by electrochemical impedance spectroscopy. Low temperature heat capacity measurements reveal a Boson peak described by a low Debye temperature of 147 K along with three independent Einstein oscillators. Overall, this indicates that soft bonding and large Ag displacements result in low-frequency optical modes that couple with acoustic modes, resulting in an extremely low thermal conductivity.

## 2. Results and discussion

### 2.1 Physical properties

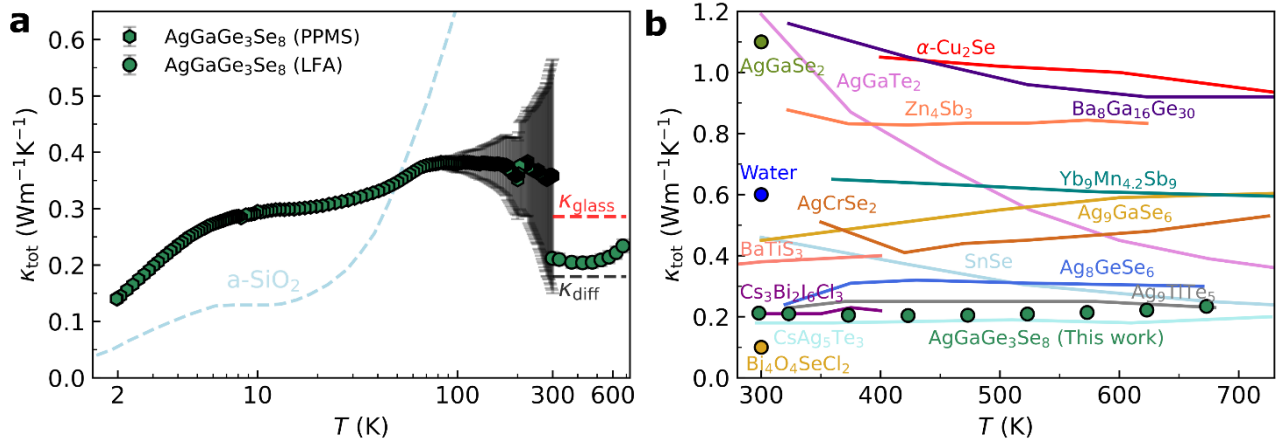

**Figure 1** a) Low temperature thermal conductivity of AgGaGe$_3$Se$_8$ measured with PPMS (green hexagons) compared with amorphous SiO$_2$ (light blue), [4] and high temperature thermal conductivity of AgGaGe$_3$Se$_8$ (green circles) measured with LFA. The calculated minimum thermal conductivity from the glass model (red dashed line) and the diffuson model (black dashed line) is included at high temperature for AgGaGe$_3$Se$_8$. b) Comparison of high temperature thermal conductivity with other ultra-low thermal conductors from the literature. [10-11, 13-24]

The low and high temperature thermal conductivity for AgGaGe$_3$Se$_8$ is shown in Figure 1a, exhibiting ultra-low thermal conductivity in the temperature range 300 K to 700 K. The thermal conductivity of AgGaGe$_3$Se$_8$ is extraordinarily low for a crystalline material, lower than amorphous SiO$_2$ (1.3 Wm$^{-1}$K$^{-1}$) and water (0.6 Wm$^{-1}$K$^{-1}$) at room temperature, and lower than many thermoelectric materials, such as the thermoelectric superionic conductors of Cu$_2$Se, Zn$_4$Sb$_3$, and Ag$_9$GaSe$_6$ (see Figure 1b). The lattice thermal conductivity is similar to the lowest thermal conductivity reported for any crystalline materials to date, such as BaTiS$_3$ (0.39 Wm$^{-1}$K$^{-1}$),[24] Tl$_3$VSe$_4$ (0.3 Wm$^{-1}$K$^{-1}$),[25] Cs$_3$Bi$_2$I$_6$Cl$_3$ (0.2 Wm$^{-1}$K$^{-1}$),[23] Cu$_4$TiSe$_4$ (0.19 Wm$^{-1}$K$^{-1}$),[26] and Bi$_4$O$_4$SeCl$_2$ (0.1 Wm$^{-1}$K$^{-1}$).[27] AgGaGe$_3$Se$_8$ is an electrical insulator, and thus the total thermal conductivity can be equated to the lattice thermal conductivity. The insulating behavior of AgGaGe$_3$Se$_8$ is apparent from its large direct and indirect band gaps of 2.17 eV and 1.91 eV, respectively, as determined by UV-VIS measurement and Tauc analysis (Figure S1).

At high temperature, the thermal conductivity of a crystalline material typically decreases with temperature, following a $\kappa \sim T^{-1}$ trend, as a result of increasing phonon anharmonicity that leads to phonon-phonon Umklapp scattering as the major scattering mechanism. At low temperatures the phonon-phonon Umklapp scattering decreases and eventually the phonon mean free path approaches the length of the crystallite, and the thermal conductivity becomes dominated by the heat capacity, resulting in the thermal conductivity following a $\kappa \sim T^3$ trend.[28] The intermediate between these two regions results in a peak-shaped temperature profile, which is referred to as the crystalline peak, often observed in the region 10-50 K. However, for AgGaGe$_3$Se$_8$ this crystalline peak is absent, and the behavior of the thermal conductivity is reminiscent that of amorphous silica, with an initial sharp increase, a plateau near 10 K, followed by a further increase and then a very weak temperature dependence from room temperature and up.[4] Between 100 K and 300 K the thermal conductivity of AgGaGe$_3$Se$_8$ becomes very flat, and the room temperature thermal conductivity from the PPMS does not match the LFA measurement. The fact that these do not match is likely due to the increased uncertainty above 150 K in the PPMS measurements due to radiation losses, and thus the LFA measurements are generally more reliable from room temperature and up.[29]

The flat to slightly increasing thermal conductivity with temperature observed for AgGaGe$_3$Se$_8$ between 300 K and 700 K is atypical for crystalline materials and resembles the behavior seen in glasses.[30] A flat temperature dependence of the thermal conductivity at high temperatures is also seen for ion conducting materials, such as Zn$_4$Sb$_3$[31] and MgAgSb,[32] and in materials with diffuson-mediated transport.[8] In a crystal the heat is transported by elastic waves using the phonon description of correlated atomic motion. However, the description of elastic waves cannot be used to describe the thermal conductivity in amorphous materials and is better described by the proposed model of Cahill *et al.*,[33] in which heat is transmitted in a random walk from one atom to its nearest and next-nearest neighbors, with a jump time equal to half a period of their oscillation. This was shown using a slightly modified version of the Einstein model, which assumes uncorrelated, Einstein oscillators. Several disordered crystalline compounds have been reported to exhibit glass-like behavior of the thermal conductivity such as $YB_{66}$,[34] $(KBr)_{1-x}(KCN)_x$,[4] $Sr_8Ga_{16}Ge_{30}$,[35-36] and more recently $Cu_{12}Sb_2Te_2S_{13}$,[37] $BaTiS_3$,[24] $Cs_3Bi_2I_6Cl_3$,[23] $Bi_4O_4SeCl_2$,[27] and $Ag_9GaTe_6$.[38]

The minimum thermal conductivity in the glass-like limit can be estimated using the Cahill-Pohl model [4] that in the high temperature limit ($T > \theta_D$) takes the form:[4]

$$\kappa_{\text{min,glass}} = 1.21\, n^{\frac{2}{3}} k_B \frac{1}{3}(2v_T + v_L).$$

Here $n$ is the number density of atoms (# atoms pr. unit cell / volume of unit cell), $k_B$ is the Boltzmann constant, and $v_T$ and $v_L$ the transverse and longitudinal speed of sound, respectively. The arithmetic average speed of sound ($v_S = 1/3(2v_T + v_L)$) can be estimated from the Debye temperature as $k_B \theta_D = \hbar \omega_D = \hbar(6\pi^2 n)^{1/3} v_S$.[6] The calculated glass limit to the thermal conductivity is plotted for $AgGaGe_3Se_8$ as the red dashed line in Figure 1a using the average Debye temperature obtained from heat capacity and PXRD experiments. It is seen that the measured thermal conductivity is lower than the glass limit in the whole temperature region. A thermal conductivity lower than the glass limit has been reported for the superionic compounds of $Cu_2Se$,[10] $Cu_2S$,[39] and $Cu_7PSe_6$.[40]

The proposed model of the glass limit to the thermal conductivity utilizes the phonon description of lattice vibrations and the limit can be viewed as the limit where the average phonon mean free path, $l$, approaches the interatomic spacing, $a$. The thermal conductivity can also be described by diffusive thermal transport using diffusons that are proposed as non-propagating, non-localized heat-carrying vibrations at intermediate energies, and is proposed to be a better description of atomic vibrations in materials with more atomic disorder, including large complex unit cells.[5-6] The minimum thermal conductivity using the diffuson model is very similar to, but lower than, the glass limit at high temperatures:[6]

$$\kappa_{\text{min,diff}} = 0.76\, n^{\frac{2}{3}} k_B \frac{1}{3}(2v_T + v_L).$$

The calculated diffuson limited thermal conductivity is plotted as the dashed black line in Figure 1a, indicating that $AgGaGe_3Se_8$ is approaching the diffuson limited thermal transport, with the experimental values just above this limit in the entire temperature regime.

The origin of the low thermal conductivity $AgGaGe_3Se_8$ possibly is related to weak soft bonding of Ag in the structure. Additionally, the large number of atoms in the unit cell ($N = 78$) will also lower the thermal conductivity, as a larger number of atoms leads to many optical phonon modes which exhibit low group velocities that do not contribute to the thermal conductivity.[41] The large number of atoms leads to increased phonon scattering and has been reported to lower the thermal conductivity proportional to $1/N$, as is reported to be the main factor to the low thermal conductivity observed in materials with large and complex unit cells such as $Gd_{117}Co_{56}Sn_{112}$.[42] Furthermore, it has been proposed by Deng *et al.* that the number mismatch (δ) between cations ($N_{cation}$) and anions ($N_{anion}$), $\delta = (N_{cation} - N_{anion})/N_{anion}$, is a good predictor for low lattice thermal conductivity in complex chalcogenides, as it is correlated with multiple phenomena such as large unit cells, vacancies, chemical bonding distortions, rattling atoms and lone pair effects.[43] The large number mismatch for $AgGaGe_3Se_8$ ($\delta$ = -0.375) and low thermal conductivity conform with the predictions of other complex Ag- and Cu-based chalcogenides.[43] The partial occupancy of the Ag sites (0.375) leads to intrinsic vacancies that can scatter high-frequency phonons and thus lower the lattice thermal conductivity. In other systems, the dynamic off-centering of the center atom in the tetrahedra induces ultra-low thermal conductivity, reported for $CsSnBr_3$[44] and $AgGaTe_2$.[45]

### 2.1.1 Low temperature heat capacity

Further insights into the thermal behavior can be gained from low temperature behavior of the heat capacity. The measured heat capacity at low temperature (1.8 K – 300 K) is shown in Figure 2a for $AgGaGe_3Se_8$. The heat capacity in the low temperature region (< 40 K) can be described using the Debye-Einstein model:[46]

$$\frac{C_p}{T} = \gamma + \beta T^2 + \sum_n A_n \Theta_{E_n}^2 (T^2)^{-\frac{3}{2}} \frac{e^{\frac{\Theta_{E_n}}{T}}}{\left(e^{\frac{\Theta_{E_n}}{T}} - 1\right)^2}$$

Here, the first term, γ, is the Sommerfeld constant that contains the electronic contribution. The second term is the Debye lattice contribution with $\beta = (1 - \sum_n A_n/3NR) \cdot (12\pi^4 N_A k_B/5\Theta_D^3)$ where $A_n$ is the prefactor of the *n*th Einstein oscillator, $N$ is the number of atoms per formula unit, $R$ is the gas constant, $N_A$ is Avogadro's constant, $k_B$ is the Boltzmann constant, and $\Theta_D$ is the Debye temperature. The third term represents the contribution from the Einstein oscillator modes with $\Theta_{E_n}$ being the Einstein temperature of the *n*th Einstein modes. In Figure 2b $C_p/T$ versus $T^2$ is fitted with the Debye mode and 1,2, and 3 Einstein modes, showing that the Debye model can not adequately describe the data, meaning it is not enough to only consider the long-wavelength acoustic phonon modes. In Figure 2c the low temperature region is highlighted in the plot of $C_p/T^3$ versus $T$, emphasizing that three Einstein modes are needed to describe the data below 5 K. This indicates the presence of localized low-frequency vibrations in $AgGaGe_3Se_8$.

The peak in $C_p/T^3$ at around 5 K is referred to as a Boson peak and ascribed to excess vibrational states over the Debye density of states.[21] It is known that the heat capacity in most crystalline solids is described well at low temperatures by the Debye model, which assumes only a single acoustic branch with a linear dispersion in the entire Brillouin zone, and ignores the existence of optical branches. The Boson peak is rarely observed in crystalline compounds and is a peculiar feature of lattice dynamics that normally occurs in non-crystalline and amorphous materials such as glasses that possess intrinsically low thermal conductivity.[21] However, the Boson peak has also been observed in rare cases of crystalline materials that exhibit very low intrinsic thermal conductivity such as clathrates,[47-48] skutterudites,[49] $AgGaTe_2$,[45] $AgSbTe_2$,[50] $CsSnBr_3$,[44] $Cu_3SbSe_3$,[51] and $TlInTe_2$.[52] The Boson peak signifies the presence of an excess phonon density of states and contribution to the heat capacity from low energy optical phonon modes and is a sign of strong phonon anharmonicity in the sample.[48, 50] The origin of the Boson peak is still debated, however, it has been proposed that localized vibrational modes, such as the rattler-modes in clathrates and skutterudites, will induce a Boson peak,[53-54] and that the intrinsic rattling vibrations of weakly bonded constituents will induce anharmonic soft optical modes that can couple with the acoustic phonons.[55]

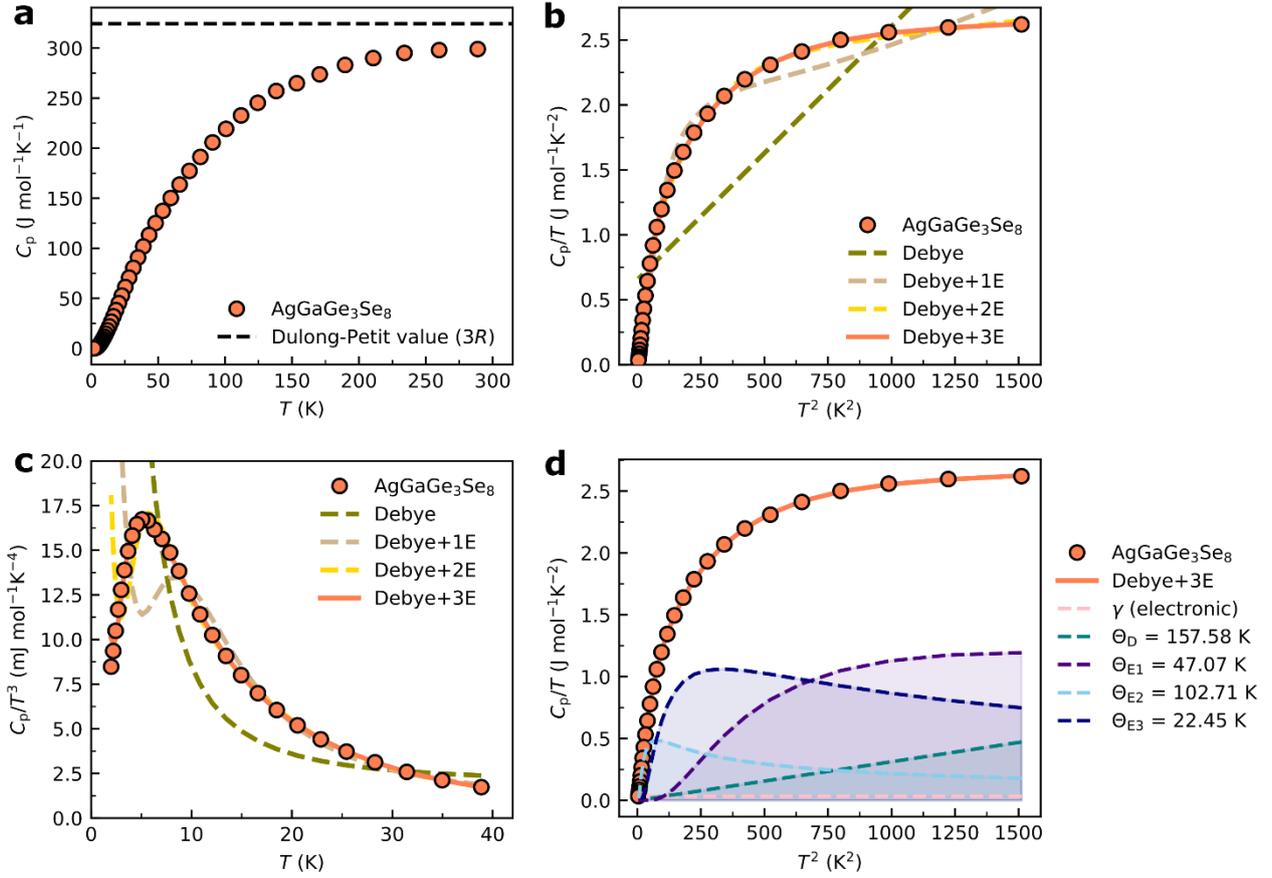

**Figure 2** Low temperature heat capacity of AgGaGe$_3$Se$_8$. a) $C_p$ as a function of temperature. b) $C_p/T$ as a function of $T^2$ with Debye-Einstein fits. c) $C_p/T^3$ as a function of $T$ with a clear Boson peak at 5 K. d) The best fit using Debye plus three Einstein modes decomposed into the contribution from the various terms.

It is apparent that three Einstein modes are needed to describe the low temperature behavior of the heat capacity of AgGaGe$_3$Se$_8$, with a resulting Debye temperature of 158 K and three Einstein temperatures of 22 K, 47 K, and 103 K (Figure 2d). The very low Debye temperature indicates soft chemical bonds, comparable with values reported for Cu$_7$PS$_6$,[56] CuIn$_7$Se$_{11}$[57] and Ag$_9$GaSe$_6$.[58] The electronic contribution is negligible which corroborates with the electrical insulating behavior. Only two Einstein modes are generally needed to describe the heat capacity in the clathrates,[48] the AgGaTe$_2$ system,[21, 45] and Ag$_8$GeSe$_6$,[59] however three Einstein modes are reported necessary for other systems such as Ag$_8$SnSe$_6$,[60] AgSbTe$_2$,[50] Cu$_7$PS$_6$,[56] Cs$_3$Bi$_2$I$_6$Cl$_3$,[23] and TlInTe$_2$.[52]

As heat is mainly transported by low frequency (long wavelength) acoustic modes, the low thermal conductivity is likely limited by a coupling to the low-frequency Einstein oscillator that inhibits heat transport. Low-frequency optical modes are often seen in materials with a host lattice containing guest atoms but can also be a significant factor for materials with a dynamic disorder of ions, as this also gives rise to low-energy optical modes.[61]

## 2.2 Powder X-ray diffraction

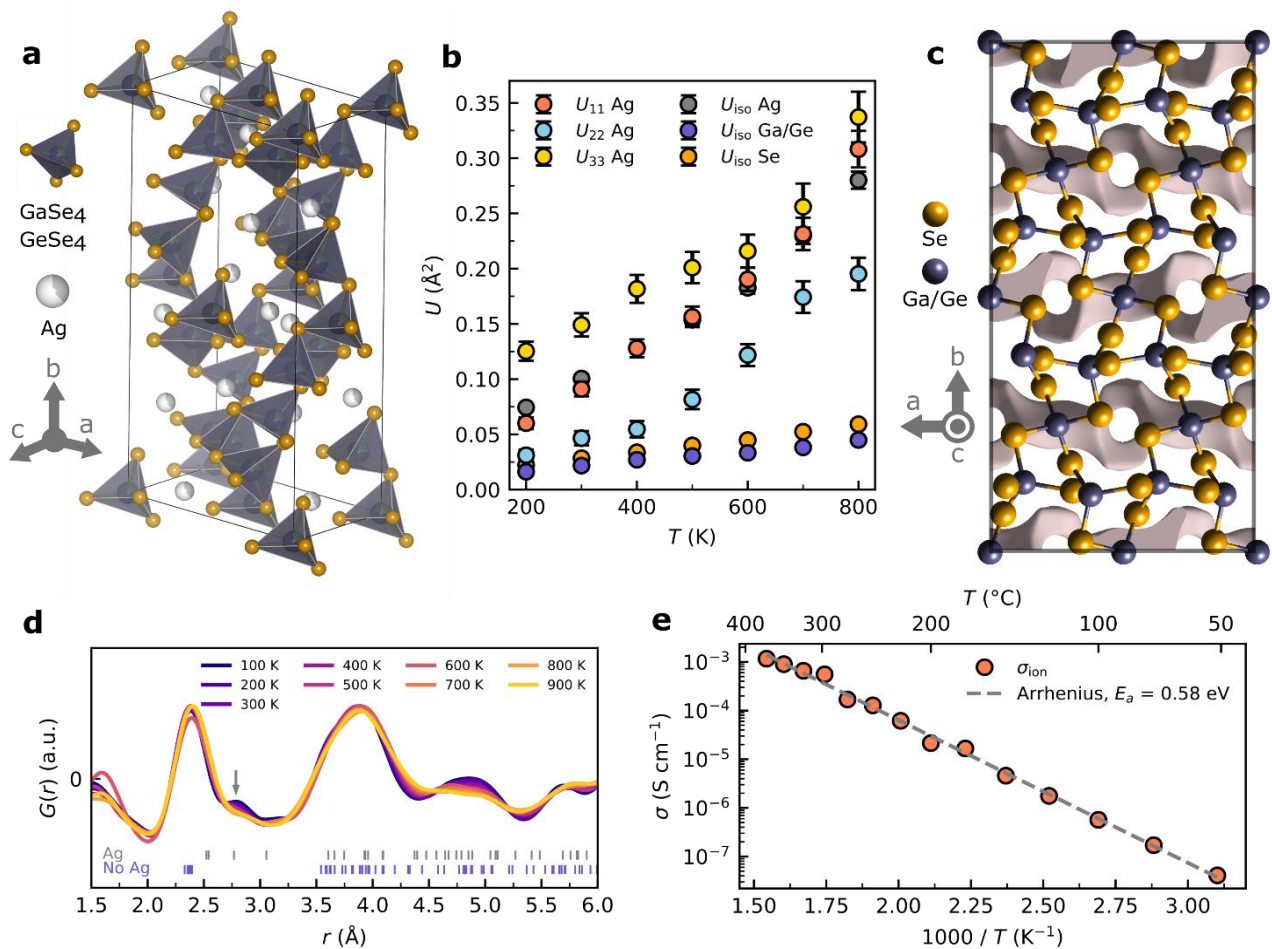

**Figure 3** a) The orthorhombic unit cell of AgGaGe$_3$Se$_8$ with Ag in grey, Ga/Ge in purple, and Se in orange. b) Refined anisotropic ADPs for the Ag site ($U_{11}$, $U_{22}$, and $U_{33}$) together with the isotropic values for the Ag site (grey) and the average values for the Ga/Ge (purple) and Se (orange) sites. c) Void space visualization projected along the *c*-axis. Generated with CrystalExplorer using an isosurface of 0.002 au. d) Pair distribution function in the short range up to 6 Å. Bond distances including Ag are marked in grey and bond distances excluding Ag are marked in purple. The grey arrow at 2.7 Å indicates the relative disappearance of a Ag-Se correlation with increasing temperature. e) Ion conductivity from electrochemical impedance spectroscopy measurement (orange) as a function of inverse temperature with the resulting Arrhenius fit (grey).

The structural origin of the extremely low thermal conductivity in AgGaGe$_3$Se$_8$ is investigated by high quality synchrotron powder X-ray diffraction. PXRD experiments are performed on as-synthesized powder before densification with spark plasma sintering (SPS) from 100 K to 900 K (Figure S2) and on powder obtained after SPS from 200 K to 900 K (Figure S3). AgGaGe$_3$Se$_8$ exhibits no phase transition in the range 100 K to 900 K and all diffraction peaks can be assigned to the orthorhombic unit cell reported by Reshak *et al.*[62] (space group *Fdd2*, unit cell shown in Figure 3a). The structural refinements are in better agreement from the sample after SPS than before SPS, as shown from selected Rietveld refinements in Figure S4. Neither of the samples show any impurity phases, but differ in the relative intensity of some Bragg reflections, similar to the discrepancy observed in PXRD patterns reported by Liu *et al.*[12] and Parasyuk *et al.*[63] The origin of this structural discrepancy is not understood, however the effect of high pressure and large currents during SPS has likely driven the Ag ions into slightly different positions, which does not occur by a simple annealing, as the sample before SPS shows the same PXRD pattern before and after heating to 900 K (Figure S4a,b). Additionally, the sample before SPS shows a significantly larger diffuse background than the sample after SPS, which cannot be described solely by the background measurement of an identical empty capillary, and also increases with temperature (Figure S5). The increasing background intensity is likely linked to diffuse scattering originating from increasing disorder in the structure with elevated temperature. The difference observed in local structure before and after SPS could be an effect of changing ion migration pathways under the influence of an electrical field, as recently reported for Cu$_2$Se under operating conditions using the maximum entropy method.[64]

From here on the refined parameters presented are from the sample after SPS due to better structural agreement. The refined unit cell parameters as a function of temperature are given in Figure S6. The refined anisotropic ADPs for Ag together with the average isotropic ADPs for the Ag, Ga/Ge, and Se sites as a function of temperature are shown in Figure 3b (ADP for all sites are shown in Figure 4 and Figure S2b). The Ag atoms exhibit a much larger ADP than the other atoms at all temperatures and it increases faster with temperature as well. The refined $U_{iso}$'s for Ag exhibit unphysically large values of 0.074 Å$^2$ at 200 K to 0.303 Å$^2$ at 800 K, which cannot solely be attributed to thermal displacements.

Refined anisotropic ADP values of Ag reveal much larger displacement along *a*- and *c*-axis ($U_{11}$ and $U_{33}$) than along the *b*-axis ($U_{22}$), indicating easier movement of Ag in the *a-c* plane than along the *b* direction. The anisotropic movement of Ag is corroborated by void space analysis where Ag is removed from the AgGaGe$_3$Se$_8$ unit cell, revealing possible ion migration pathways in the structure.[65] Calculated void spaces in the AgGaGe$_3$Se$_8$ structure are visualized along the *c*-axis in Figure 3c, and along the *a*- and *b*-axis in Figure S7. Clear channels are revealed in the structure that are connected both along the *a*- and *c*-axis, but not along the *b*-axis. Channels in the *a-c* plane occur at an isovalue of 0.0016 au, while channels open along the *b*-axis at 0.0039 au, isovalues that are comparable with common Li-ion conductors.[65] This agrees well with the observed enlarged ADPs for Ag along the *a*- and *c*-axis. The unphysical large ADPs for Ag indicate that the current model for the structure does not correctly account for the atomic position of Ag in the structure. The Ag-atoms are likely more disordered in the *a-c* plane. The partly occupied Ag sites will also lead to differences in the local environment and relaxation of the surrounding atoms. Relaxation of nearest-neighbor atoms to vacancies can be quantified based on modelling of diffuse scattering as shown for thermoelectric half-heusler[66] or InTe materials,[67] as well as yttria-stabilized zirconia.[68] Due to the large amount of vacant Ag positions in the structure a similar effect is expected for $AgGaGe_3Se_8$, which is possibly the origin of the differences observed in the relative intensities. Another effect could be that the stoichiometry assumed in the Rietveld model is incorrect, however compositional analysis by EDX and XRF (Table S1) reveals the stoichiometry to be close to the nominal stoichiometry, and is rather slightly deficient in Ge and Se. The Rietveld analysis shows that further investigations of the crystal structure using single crystal analysis are needed to fully understand the atomic arrangement in $AgGaGe_3Se_8$. Elucidation of the true local structure of the disordered Ag atoms may be done using the 3D-ΔPDF method based on single crystal X-ray,[69-70] neutron,[71] or electron scattering.[72]

**2.3 Pair distribution function analysis**

The calculated pair distribution function (PDF) of $AgGaGe_3Se_8$ from 100 K to 900 K is plotted in Figure 3d in the short range regime (1.5 Å to 6 Å). The complex unit cell results in a multitude of overlapping interatomic distances above 3.5 Å, however the local bond environments of the Ga/Ge-$Se_4$ tetrahedra (around 2.33-2.38 Å) and the Ag-$Se_4$ tetrahedron (around 2.52-3.05 Å) can be separated in space. The first peak around 2.3 Å only has contribution from Ga/Ge-Se distances, the second peak at around 2.7 Å (indicated by grey arrow) only has contribution from one Ag-Se distance. The peak with Ag-Se correlations exhibit a much stronger decrease with temperature than the non-Ag containing correlations. This suggest that the Ag atoms gradually lose coordination to the neighboring Se atoms with temperature and becomes more disordered, and could be associated with an increased Ag mobility as reported in a similar analysis for the Ag ion conductor of $AgCrSe_2$.[73]

The increasing Ag disorder with temperature is accompanied by an increasing bond distance between the Ag atom and the surrounding Se atoms that increases from 2.71 Å at 200 K to 2.76 Å at 800 K (Figure S8a). The substantial expansion of the Ag tetrahedron correlates well with the largest increase in the ADPs. In contrast, the average Ga1/Ge2-Se and Ga2/Ge2-Se bond distances are nearly constant with temperature (Figure S8a). As a whole, the unit cell expands anisotropically with temperature, with the largest expansion along the *a*- and *b*-axis and lowest along the *c*-axis (Figure S8b).

**2.4 Ion conductivity from electrochemical impedance spectroscopy**

Evidence of Ag ion conduction in the structure is obtained by electrochemical impedance spectroscopy measurements, performed on $AgGaGe_3Se_8$ up to 400 °C in the frequency range of 0.11 Hz to 11 MHz. The absolute impedance decreases with temperature and the phase approaches a constant value of around -20° in a large frequency region (0.1 Hz to 0.1 MHz) after 300 °C (Figure S9). This is accompanied by the emergence of a low frequency semicircle and high frequency mass-transfer limited tail in the Nyquist plots (Figure S10), which is a behavior typically observed for ion conduction in solid-state ion conductors. The extracted ion conductivities with temperatures are shown in Figure 3e, exhibiting a monotic increase in conductivity with temperature with no discernible transition temperature (an example of EIS fit at 400 °C is shown in Figure S11). Fitting with an Arrhenius equation the activation energy is calculated to $E_a = 0.58$ eV, which is comparable to other insulating materials exhibiting Ag ion conduction such as $AgGaS_2$ ($E_a = 0.62$ eV),[74] while being substantially lower than the reported activation energy of the superionic silver conductors of AgI ($E_a = 0.10$ eV)[75] and $AgCrSe_2$ ($E_a = 0.11$ eV).[76]

Materials exhibiting ion conduction are usually correlated with low thermal conductivity, however the magnitude of the ion conduction has recently been reported to not directly influence the thermal conductivity, where a difference in $Ag^+$ ion conductivity in $Ag_8Ge_{1-x}Sn_xSe_6$ argyrodite compounds over several orders of magnitude does not change the lattice thermal conductivity.[8] Thus, the low thermal conductivity cannot be directly explained by a 'liquid-like' sublattice of mobile ions but can instead be attributed other effects such as atomic disorder, anharmonicity, low-frequency optical modes due to soft bonding, and in general complex crystal structures. Strikingly, the very low thermal conductivity in the superionic conductor of $Zn_4Sb_3$ is often attributed to a disordered structure consisting of Zn interstitials even though the low thermal conductivity of the disordered $Zn_4Sb_3$ can still be reproduced in the fully ordered $Zn_6Sb_5$ structure,[77] and strong anharmonicity is found in the weak multicenter bonding inside the $Zn_2Sb_2$-rhombi of the ZnSb phase even at 20 K.[78]

## 2.5 Debye and Einstein analysis from PXRD

The evolution of the isotropic ADPs ($U_{iso}$) with temperature can be directly related to the Debye temperature of the lattice by[79-80]

$$U_{iso} = \frac{3\hbar^2 T}{mk_B \Theta_D^2}\left(\frac{T}{\Theta_D}\int_0^{\Theta_D/T}\frac{x}{e^x - 1}dx + \frac{\Theta_D}{4T}\right) + d^2$$

Here, $m$ is the average mass, $\hbar$ is the reduced Planck's constant, and $d$ is an empirical term describing temperature independent disorder. Fits up from 200 K to 600 K of the measured $U_{iso}$ for each site and the mass averaged $U_{iso}$ for the average structure of $AgGaGe_3Se_8$ are shown in Figure 4a. The Ag atoms exhibit the substantially lowest Debye temperature ($\Theta_D$ = 71 K). The three Se sites have Debye temperatures in the range 167 K to 193 K while the Ga and Ge sites are between 207 K to 220 K. This indicates that the interconnected $GaSe_4$ and $GeSe_4$ tetrahedra forms a rigid framework whereas the Ag atoms are more diffuse and disordered in the structure and could exhibit rattling or migration behavior. The shortest Ag-Se bond is between Ag and Se2, which explains the lower Debye temperature and higher disorder parameter for the Se2 site compared to the other Se sites. The mass averaged Debye temperature determined ($\Theta_{D,PXRD}$ = 156 K) agrees well with the value obtained from the heat capacity measurements ($\Theta_{D,C_P}$ = 158 K). If the Ag atoms act as rattlers they are better described by independent Einstein oscillators, where the interactions with the surrounding network of atoms are neglected and the resulting phonon dispersion is flat. A similar model can successfully describe the independent rattler modes of guest atoms residing inside a host framework of cage structures in the clathrate compounds.[80-82]

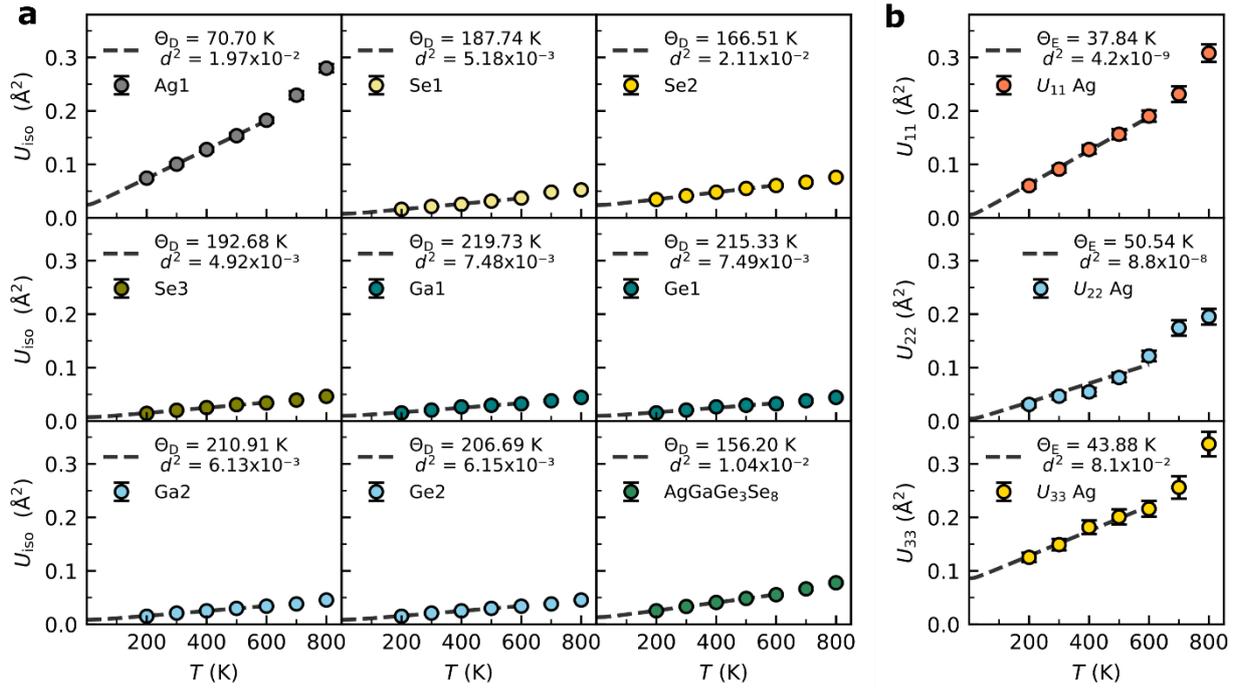

**Figure 4** a) Refined isotropic ADPs from 200 K to 800 K for all the sites in the AgGaGe$_3$Se$_8$ structure with corresponding Debye fits up to 600 K (blacked dashed lines) and the resulting Debye temperature and disorder parameter (*d*) for each site along with the averaged Debye temperature for the AgGaGe$_3$Se$_8$ structure (green). b) Einstein modelling with the associated Einstein temperature of the anisotropic atomic displacement parameters for Ag.

In the Einstein model the temperature dependence of the ADPS can be described by[80, 82]

$$U_{ij} = \frac{\hbar}{2mk_B\Theta_{E,ij}} \coth\left(\frac{\Theta_{E,ij}}{2T}\right) + d^2$$

Where $\Theta_{E,ij}$ is the Einstein temperature corresponding to the ADP tensor element $U_{ij}$ and $d^2$ is again used to describe temperature independent disorder. The Einstein model is applied to the isotropic and anisotropic ADPs of Ag (Figure 4b), resulting in a very low average Einstein temperature ($\Theta_{E,iso}$ = 41 K), and lowest Einstein temperature for $U_{11}$ along the *a*-axis of 38 K. The Einstein temperature is lower than, and the disorder parameter is similar to, usual values for guest rattler atoms in clathrates structures,[80-82] which suggest strong disorder and weak bonding of Ag in the crystal structure of AgGaGe$_3$Se$_8$. The average Einstein temperature corresponds to an average Einstein frequency of 0.85 THz, which is comparable, but lower, than the Einstein frequency of 1.2 THz reported for Ag in the Ag-ion conductor Ag$_8$GeSe$_6$.[8] The number of Einstein modes is usually connected to the dimensionality of the atomic vibrations. In the clathrates the rattling of the two guest atom sites, and in- and out-of-plane vibrational motion of the guest atom in the large cage, is characterized by their own Einstein modes.[83] In the compound of AgGaGe$_3$Se$_8$ the three Einstein modes observed in both heat capacity and PXRD analysis could be linked to the three independent vibrational directions of Ag.

## 3. Conclusion

In summary, we report on the ultra-low glass-like thermal conductivity observed in the quarternary chalcogenide AgGaGe$_3$Se$_8$ of only 0.2 Wm$^{-1}$K$^{-1}$ at room temperature. The origin of the extremely low thermal conductivity is linked to structural anharmonicity as evident from a Boson peak in the heat capacity around 5 K, which needs a Debye mode and three Einstein modes to be adequately described. Multi-temperature PXRD data reveal enormous anisotropic ADPs for the Ag atoms that exhibit the largest vibrations in the *a-c* plane, corroborating measurements of ionic conductivity at elevated temperatures. The low Debye temperature obtained from both heat capacity measurement and PXRD analysis ($\Theta_{D,Cp} = 158$ K and $\Theta_{D,PXRD} = 156$ K) sediments the presence of a soft lattice. These findings contribute to a better understand of the peculiar glass-like thermal conductivity observed in crystalline materials.

## 4. Experimental methods

*Polycrystalline synthesis:* For the polycrystalline synthesis of AgGaGe$_3$Se$_8$ stoichiometric amounts of Ag, Ga, Ge, and Se were weighed (total of 5 g) and loaded into a quartz ampoule. The quartz ampoule was evacuated to <10$^{-4}$ bar and closed using a propane-oxygen flame. The synthesis was carried out in a two-zone furnace. Initially both zones were kept just above the melting point of selenium at 498 K for 10 h and subsequently equilibrated just below the boiling point of selenium at 923 K, to ensure consumption of most of the selenium. Then, the hot zone was heated to 1223 K while the cold zone was held at 673 K. Under these conditions, a heterogeneous reaction will take place in the hot zone between liquid components of Ag, Ga, Ge, and GeSe$_2$, and gaseous components of Se and GeSe$_2$. Condensation of the latter will occur in the cold zone, which helps reduce the rising pressure inside the ampoule. After 15 h at this temperature gradient most of the gaseous components were expected to have reached the hot zone, and the temperature was slowly increased in the cold zone to 1273K at 1K min$^{-1}$ to consume the rest. After 20 h at the maximum temperature the mixture was expected to be homogeneous, and the furnace was slowly cooled at 0.5 K min$^{-1}$ to 723K and finally to room temperature. For physical property measurements a pellet was densified by spark plasma sintering. The obtained polycrystalline ingot was ground and sieved to particle sizes <63 μm, loaded into a graphite die with a diameter of 1.27 cm and were sintered for 10 min at 530 °C and 50 MPa using an SPS-515 instrument (SPS Syntex Inc., Japan).

*Synchrotron powder X-ray diffraction:* Synchrotron PXRD experiments were performed at BL44B2 beamline, Spring-8, Japan. The data were collected on a the MYTHEN microstrip detector system (OHGI) [84] in transmission geometry using a wavelength of λ = 0.569758 Å. Polycrystalline powder of AgGaGe$_3$Se$_8$ was sieved to particles sizes <63 μm and packed in quartz capillaries with inner diameter of 0.2 mm to ensure a suitably low absorption. Calibrated temperatures between 100 K and 900 K were reached using a combination of a heat gun and a cryogenic stream. The synchrotron PXRD data obtained at BL44B2 exhibit simultaneous high angular resolution and high measured $q$-range ($Q_{max}$ ~ 21 Å), which allows for dual-space structural analysis from a single dataset, *i.e.* Rietveld analysis in reciprocal space combined with pair distribution function (PDF) analysis in real space.[85] Void space analysis[86] was performed in CrystalExplorer[87] by systematically varying the isovalue until connected channels appeared.

*High-temperature thermal conductivity measurement:* Thermal diffusivity, $D$, was measured using the laser flash method (LFA) with a Netzsch LFA 457. Thermal conductivity, κ, was calculated from $\kappa = DdC_p$, where $d$ is the thickness of the pellet and $C_p$ is the heat capacity. $C_p$ was measured against a reference sample (Pyroceram 9606) of known density and heat capacity. The AgGaGe$_3$Se$_8$ sample used had a thickness of $d$ = 1.882 mm and density of $\rho$ = 4.703 g/cm$^3$.

*Low-temperature physical property measurements:* Low temperature heat capacity and thermal conductivity was measured on a physical property measurement system (PPMS, Quantum Design, US) from 2 K to 300 K using the Thermal Transport Option (TTO) under high vacuum. Following sample dimensions were measured on the AgGaGe$_3$Se$_8$ sample, cross-sectional area of 3.381 mm$^2$ and distance between probes of 3.85 mm.

***UV-VIS measurements:*** UV-VIS measurements were performed on a Shimadzu UV-3600 instrument. The Kubelka-Munk function was calculated from the diffuse reflectance, $R_\infty$, as $F(R) = (1-R_\infty)^2/2R_\infty$. The band gap is then extracted using the Tauc method, by replacing the energy-dependent absorption coefficient, α, with $R_\infty$, in the Tauc equation $(\alpha h\nu)^{1/\gamma} = (F(R_\infty)h\nu)^{1/\gamma} = B(h\nu - E_g)$, where $h$ is the Planck constant, $\nu$ is the frequency of the photon, $E_g$ is the band gap and B is a constant. The γ is a factor depending on the electronic transition and is equal to 1/2 in a direct band gap and 2 in an indirect band gap.

***Electrochemical Impedance Spectroscopy measurements:*** The electrochemical impedance spectroscopy (EIS) measurements were performed from 0.11 Hz to 1.1 MHz on a Biologic MTZ-35 impedance analyzer with a molybdenum sample holder equipped with a 4-point setup to eliminate inductance, resistance, and capacitance from the setup. The experiment was performed on a square sample with a side length of 3.92 mm and a thickness of 1.24 mm. Data were collected from room temperature to 400 °C in Ar atmosphere. The data was fitted using an equivalent circuit of R1-p(CPE1,R2-CPE2), where R1 represents offset resistance, R2 represent the charge transfer resistance, and CPE1 and CPE2 are constant phase elements representing the double layer capacitance and the mass transfer, respectively. In the equivalent circuit, elements in parallel are represented by 'p' and elements in series by a dash '-'. The ionic conductivity is calculated as d/(A·R2), where R2 is the charge transfer resistance, d is the thickness of the sample, and A is the area.